\documentstyle[epsf,preprint,eqsecnum,floats,prd,aps]{revtex}

\begin{document}
\draft

\title{Self-similar scalar field collapse: naked
singularities and critical behaviour}

\author{Patrick R Brady }
\address{
Department of Physics,  The University,  Newcastle upon Tyne
NE1 7RU}
\preprint{NCL94-TP12}

\maketitle
\date{\today}

\begin{abstract}
\widetext

Homothetic scalar field collapse is considered in this article.  By making a
suitable choice of variables the equations are reduced to an autonomous
system. Then using a combination of numerical and analytic techniques
it is shown that there are two classes of solutions.  The first
consists of solutions with a non-singular origin in which the scalar
field collapses and disperses again.  There is a singularity at one
point of these solutions,  however it is not visible to observers
at finite radius.  The second class of solutions includes
both black holes and naked singularities with a critical evolution
(which is neither) interpolating between these two extremes.  The properties
of these solutions are discussed in detail.  The paper also contains some
speculation about the significance of self-similarity in recent numerical
studies.
\end{abstract}

\clearpage

\narrowtext


%
%
%
%

\section{Introduction}

That gravitational collapse leads to black hole formation is widely
accepted,  yet comparitively little is known about the generic
features of collapse.  This is well exemplified by the lack of a
precise formulation of the cosmic censorship hypothesis\cite{cosmicc}.
 The reason for this is simply our lack of the mathematical tools
necessary to analyse the evolution of generic initial data, although some
progress is being made on this front~\cite{Christ&Klain}. Moreover the
complexity of the Einstein field equations counsels retreat to a more
tractable system which may capture the essence of the physics.
Spherically symmetric general relativity coupled to a massless scalar
field is one such system.  Indeed this model has been studied in
great detail both analytically~\cite{Christ1}  and
numerically~\cite{numeric,Goldpir}.  Christodoulou has rigorously
established global existence and uniqueness of solutions to the
Einstein-scalar field equations and has discussed the general
properties of these solutions~\cite{Christ1}. He has even established
a sufficient condition for the formation of a trapped surface in the future
evolution of a given initial data set~\cite{Christ2}.   Numerical
investigations have also provided useful insights into black hole
formation. The most recent of these studies,  by Choptuik~\cite{Chop},
has revealed several intriguing phenomena which were hitherto unknown.

Choptuik considered the numerical evolution of initial data sets
characterised by a single parameter ($p$ say). The resulting families
of solutions, ${\cal S}[p]$, include both geometries containing black
holes, and geometries with only slight deviations from flatness,
depending on the value of $p$.  Between these two extremes lies a
critical evolution with $p=p^*$,  which signals the transition between
complete dispersal and black hole formation.  His most interesting
results pertain to near critical,  $p \simeq p^*$, evolutions which
exhibit a particularly simple strong field behaviour.  In fact,  his
results strongly suggest that near critical evolutions may be described by a
single, universal solution of the field equations.  Two quantitative
features also emerge from his work: (i) Near critical evolutions
contain echoes in the strong field region (i.e.  form
invariant quantities exhibit a scaling relation like (\ref{echoes})
below,  with $\Delta \simeq -3\cdot 4$).  Based on his extensive investigations
Choptuik has conjectured that the exactly critical solution has an infinite
train
of echoes in the strong field region approaching the singularity.
(ii)  When black holes form in near critical evolutions the mass
scales as $M_{bh} \sim |p -
p^*|^{\beta}$,  where $\beta$ is an
apparently universal constant determined numerically to be about
$\cdot 37$

Obtaining analytical results about solutions which exhibit a discrete
self-similarity of the type discussed by Choptuik has proved to be extremely
difficult~\cite{Viq}.  In~\cite{Japs} it has been argued that the piling up of
the echoes as the singularity is approached suggests that a continuous
self-similarity may be a good approximation in this neighbourhood.   Therefore
it seems worthwhile to examine collapse under the
assumption of a continuous self-similarity.  Furthermore I would like
to draw attention to a possible explanation of the echoing,
discovered by Choptuik, in terms of a continuous self-similarity.
In curvature coordinates the spherical line element may be written as
	\begin{equation}
	ds^2 = -\alpha^2 dt^2 + a^{2} dr^2 + r^2 d\Omega^2
	\end{equation}
where $\alpha$  and $a$ are functions of both $r$ and $t$.
Self-similarity implies that there exists a time coordinate such that
the metric depends only on the combination $x = r/t$, consequently
quantities like $a(r,t)$ satisfy the relation
	\begin{equation}
	a(e^{\Delta} r, e^{\Delta} t)  =  a(r, t) \label{echoes}
	\end{equation}
for arbitrary values of $\Delta$.  Now,  consider a coordinate
tranformation to a new time, $T$ say, in terms of which $t= f(T)$
where the function $f$ satisfies
	\begin{equation}
	f( e^{\Delta} T )  =  e^{\Delta} f(T)
	\end{equation}
only for discrete values of $\Delta$.  In terms of this new time
coordinate the form invariant quantities examined by Choptuik,  along
with $a(r/f(T))$, would all appear to have a discrete (rather than
continuous) self similarity.  How could this be the origin of the
discreteness in the numerical work, after all Choptuik~\cite{Chop}
states quite clearly that he uses central proper time to describe the
solutions -- this corresponds to the similarity time $t$ in a spacetime which
is {\em everywhere} self-similar.
However it is likely,  and indeed suggested by numerical work, that
(discrete) self-similarity holds
only on  intermediate scales, $0 < r_1 \le r \le r_2$ say.  (This has
been further highlighted in
numerical studies of radiation fluids~\cite{Evans}.)   Thus  it is
possible that the time coordinate used in the numerical studies does {\em not}
correspond to the
similarity time.  It should be noted that even if this is true the
appearance of the universal scaling constant $\Delta \simeq  3\cdot 4$
continues to be a mystery.

In any case, this paper is concerned with homothetic
spacetimes;  it is assumed that there exists a
vector field $\xi$ such that the spacetime metric satisfies ${\cal
L}_{\xi} g = 2 g$, where ${\cal L}_{\xi}$ denotes the Lie derivative
with respect to $\xi$.  The homothetic collapse of perfect fluids has
received a great deal of attention during the past few
years~\cite{homothetic}, where the main thrust of work was searches for
naked singularities~\cite{Naked}.  The problem of homothetic, scalar
field collapse has received comparitively little
attention~\cite{Goldpir},  and no examples of naked singularities which evolved
from regular initial data
were previously known for this form of matter.

The reduction of the scalar field equations to an autonomous system allows me
to give a detailed description of all the solutions.  Two interesting features
emerge from the analysis:  Firstly the existence of solutions with naked
singularities,
and secondly the occurence of phase transitions in the system. The critical
point
behaviour is of two distinct forms.  The first is a phase transition from
solutions which (roughly) represent dispersal, to geometries representing black
holes.   It has been suggested elsewhere~\cite{Eardley} that the critical
solution implied by
the numerical results may be both on the verge of black hole
formation and being a naked singularity.  We will see below that the
second phase transition corresponds to such a situation --
the critical evolution lies at the boundary between black holes and
naked singularities.
In this article the term black hole is used rather loosely to mean that an
apparent horizon exists and precedes a central singularity. Of course it  is
possible to obtain an asymptotically flat solution to the Einstein-scalar
equations by cutting off the
self-similar evolution at some advanced time and matching it to a
less symmetric (not self-similar) exterior,  in this way one would obtain an
asymptotically flat, black-hole spacetime.

The paper is arranged as follows:  In
section~{II} the
field equations
for the self-similar collapse are derived in terms of a retarded time
coordinate $u$ and a radial coordinate $r$.  The most general (see appendix A)
scalar
field evolution consistent with the homothetic symmetry of the
spacetime is $\psi = \overline{h} (r/ |u|) - \kappa \ln |u|$,  where $\kappa$
is an arbitrary (positive) constant.    These equations have already
been derived in~\cite{Goldpir} and used to provide initial data in a
search for  naked singularities.  A field redefinition transforms the
equations into a non-linear autonomous system which is amenable to
standard techniques~\cite{Nonlindiff}. It is then shown that the solutions fall
into two distinct
classes depending on the value of $\kappa$.  Solutions of the first type
($\kappa \ge 1/\sqrt{4\pi}$) are
shown  to be singular only at one point on
$r=0$.   This singularity is not naked;  distant observers cannot see
it by waiting a finite proper time.   The transition point is at $\kappa =
1/\sqrt{4\pi}$,
although the
exactly critical solution actually belongs to the first class.

 For each $\kappa < 1/\sqrt{4\pi}$ there exists a continuous infinity of
solutions with a non-singular origin.  For a given value of $\kappa$
solutions can represent black holes or naked singularities,  with a critical
point evolution interpolating between these two extremes.   The analytic
examples of phase transitions which have been discussed in the
literature~\cite{Japs} are of this type,  in fact they correspond to
the case $\kappa = 0$ when the equations can be integrated
in closed form.  This solution is briefly discussed to explain how it
fits into the general analysis presented here.
Section~{IV} contains some discussion of the results,
and their possible significance.

The notation of~\cite{MTW} is adopted throughout the paper,  and detailed
calculations
which might distract from the main line of thought are relegated to the
appendices.

%
%
%
%

\section{The field equations}

Using retarded Bondi coordinates $\{ u, r, \theta, \phi \}$ the
spherical line element may be written
	\begin{equation}
	ds^2 = - g \overline{g} du^2 - 2 g du dr + r^2 d\Omega^2
	\end{equation}
where $g = g(u,r)$, $\overline{g} = \overline{g}(u,r)$ and $d\Omega^2$ is the
standard line element on the unit two-sphere.  The origin is singular unless
$\overline{g} = g$ when $r=0$.
Furthermore it is convenient to normalise the coordinate $u$ so that
it represents proper time for an observer at the origin, thus write
	\begin{equation}
	\overline{g}(u,0) = g(u,0) = 1 \label{eqn:bc1} \; .
	\end{equation}
The Einstein-scalar field equations are then
	\begin{eqnarray}
	(\ln g)_{,r} &=& 4\pi r (\psi_{,r})^2 \; , \label{eqn:efe1}\\
	(r\overline{g})_{,r} &=& g \; , \label{eqn:efe2} \\
	(\overline{g} / g)_{,u} &=& 8 \pi r g^{-1} \left[ (\psi_{,u})^2 -
	\overline{g} \psi_{,u}
	\psi_{,r} \right] \; , \label{eqn:efe3}
	\end{eqnarray}
where a comma denotes partial differentiation.  $\psi = \psi (u,r)$ is
a massless, minimally coupled scalar field satisfying
	\begin{equation}
	(\overline{g} r^2 \psi_{,r})_{,r} = 2r \psi_{,u} + 2 r^2 \psi_{,ru}
	\; . \label{eqn:efe4}
	\end{equation}

Spherical symmetry allows the introduction of a
local mass function $m(x^{\alpha})$~\cite{P&I&Zann} defined by
	\begin{equation}
	1 - \frac{2m(x^{\alpha})}{r} := g^{\alpha\beta} r_{,\alpha}
		r_{,\beta} = \frac{\overline{g}}{g} \; ,
	\label{eqn:mass1}
	\end{equation}
where $r$ is the function which determines the
area of the two-spheres.
This mass function agrees with both the ADM and Bondi masses in the
appropriate
limits,  and is equivalent to the Hawking quasi-local mass in this
case.

\subsection{Self-similar ansatz}

The existence of a homothetic symmetry in a spherical spacetime
implies that the metric depend only on
$x = r/|u|$,  and that the scalar field evolve as
	\begin{equation}
	\psi = \overline{h}(x) - \kappa \ln |u| \label{eqn:sf}
	\end{equation}
where $\overline{h}$ is some function to be determined and $\kappa$
is constant  (see Appendix A for a
proof of this fact).   Writing
	\begin{equation}
	\overline{h} (x) = \int^x_0 \frac{\gamma(\xi)}{\xi} d\xi \; ,
	\end{equation}
the self-similar equations derived from
(\ref{eqn:efe1})-(\ref{eqn:efe3}) are
	\begin{eqnarray}
	(x \overline{g} )' &=& g \; , \label{eqn:ss1}\\
	x g' &=& 4 \pi g (\gamma)^2 \; , \label{eqn:ss2} \\
	g - \overline{g} &=& 4 \pi \left[ 2\kappa^2 x - (\overline{g} - 2x)
(\gamma^2 +
	2 \kappa \gamma)\right] \; , \label{eqn:ss4}
	\end{eqnarray}
where a prime ($'$) denotes differentiation with respect to $x$.   In
deriving (\ref{eqn:ss4}) from (\ref{eqn:efe3}) it is necessary to use
(\ref{eqn:ss1}) and
(\ref{eqn:ss2}) to eliminate derivatives of $g$ and $\overline{g}$.   The
scalar field evolution is determined by
	\begin{equation}
	x (\overline{g} - 2x) \gamma' = 2\kappa x - \gamma ( g - 2x )
	\; . \label{eqn:ss3}
	\end{equation}
It is now a straightforward matter to show that
(\ref{eqn:ss1})-(\ref{eqn:ss4}) imply (\ref{eqn:ss3}) provided $\gamma
\ne - \kappa$.  These equations have been derived
previously by Goldwirth and Piran~\cite{Goldpir} and used to provide
boundary conditions in a numerical search for naked singularities.

At the origin (\ref{eqn:bc1}) and (\ref{eqn:ss4}) imply either
$\gamma(0) = 0$ or $\gamma(0) = -2\kappa$.  Directly evaluating the
trace of the stress-energy tensor for the scalar field one finds
	\begin{equation}
	T{^{\alpha}}_{\alpha}  =
	\frac{\overline{g}}{g r^2} \left[ \gamma^2  - \frac{x}{\overline{g}}
	(\gamma^2 + \kappa \gamma)
	\right] \stackrel{r\rightarrow 0}{\longrightarrow}
	\frac{\gamma^2}{r^2}
	\: .  \label{eqn:T}
	\end{equation}
Clearly the solution can have a non-singular origin only if
	\begin{equation}
	\gamma(0) = 0 \label{eqn:bc2} \; .
	\end{equation}
This completes the specification of the initial conditions for the equations
(\ref{eqn:ss1})-(\ref{eqn:ss4}).  It is however convenient to recast the system
in an
autonomous form before discussing the solutions.

\subsection{An equivalent autonomous system}

Analysis of the above equations is facilitated by the field
redefinitions
	\begin{equation}
	y = \overline{g}/g \; , \ \ \
	{z} = x/\overline{g} \; , \label{fielddef}
	\end{equation}
and the introduction of a new coordinate
	\begin{equation}
	\xi = \ln x \; .
	\end{equation}
Upon substitution into (\ref{eqn:ss1}), (\ref{eqn:ss2}) and
(\ref{eqn:ss3}) one obtains the three dimensional, non-linear
autonomous system
	\begin{eqnarray}
	\dot{{z}} &=&  {z} \left( 2 - y^{-1}\right) \; ,
	\label{eqn:aut1} \\
	\dot{y} &=& 1 - (4 \pi \gamma^2 + 1)y \; , \label{eqn:aut2} \\
	(1-2{z}) \dot{\gamma} &=& 2\kappa{z} - \gamma (y^{-1} -
2{z})\; . \label{eqn:aut3}
	\end{eqnarray}
The system is effectively two dimensional, however, since  $\gamma$ is
determined by the algebraic relation
	\begin{equation}
	\gamma = - \kappa \pm \sqrt{ \frac{ (1 + 4\pi \kappa^2)
	-y^{-1}}{4\pi (1 - 2 {z})}} \; , \label{gamma}
	\end{equation}
provided $y \ne 1/(1 + 4\pi\kappa^2)$.  Further discussion is
therefore couched in terms of a projection into the $y{z}$-plane.
Consistent with the initial condition (\ref{eqn:bc2}) the
positive square root  is taken in (\ref{gamma}), however it must be
emphasised that solutions may still
evolve continuously onto the other
leaf of the surface defined by taking the negative square root above.
Indeed it is solutions of this type which have naked singularities.

Requiring that the mass function, defined in (\ref{eqn:mass1}), should be
positive or
zero implies $y\le 1$,  while $\gamma$ is real only if
	\begin{equation}
	\frac{ (1 + 4\pi \kappa^2) - y^{-1} }{
	 1 - 2 {z}}  \ge 0 \; .
	\end{equation}
Black hole formation is signalled by $y \rightarrow 0$
(technically this is the condition which locates an apparent horizon in
the spacetime).  The
continuation of the solution to negative values of $y$ will not be
considered in the sequel,  thus, integral curves of interest lie in the strip
$0 \le y \le 1$.

It should be noted that $\gamma$ is not continuous at
	\begin{equation}
	{z} = 1/2\; , \ \   y = 1/(1 + 4\pi \kappa^2)\; ,
	\label{eqn:singpt}
	\end{equation}
thus invalidating the usual existence and uniqueness theorems for systems of
ordinary
differential equations at this point.  This has the important consequence that
integral curves of the differential equations may intersect at
(\ref{eqn:singpt}).  Furthermore,  the continuation of such solutions is not
always uniquely defined,  in some cases there exists an infinite family of
possibilities.
This is discussed in more detail below and in appendix~C.

The character of the solutions shows a strong dependence on the value
of  $\kappa$.  Solving (\ref{eqn:ss1})-(\ref{eqn:ss3}),  subject to
the regularity conditions on $r=0$, gives rise to two essentially
different classes of solution according as $4\pi\kappa^2$ is greater
than or less than unity; they are solutions which do not contain black
holes,  and solutions which contain either black holes or naked
singularities.  The critical evolution ($4\pi\kappa^2= 1$) also belongs
to the first class. The transition from one class to the other as one adjusts
$\kappa$ is similar to the behaviour discussed by Choptuik~\cite{Chop}.   A
second type of phase transition from solutions containing black holes to those
with naked singularities occurs for {\em each} value of $\kappa$ in the range
$0< 4\pi\kappa^2 <1$.  This may simply be
an artifact of the restriction to self-similar continuations
past the point (\ref{eqn:singpt}) however no other continuations are
considered here.   A similar phenomenon has been observed in Tolman-Bondi
collapse~\cite{Joshi}.

The system (\ref{eqn:aut1})-(\ref{eqn:aut3}) has two stationary points,  one on
either leaf of (\ref{gamma}), given by
	\begin{equation}
	y_{\pm} = \frac{1}{2}\; , \ \  {z}_{\pm} = \frac{1}{1 \pm \sqrt{4\pi}
	\kappa} \; , \ \
	\gamma_{\pm} = \pm \frac{1}{\sqrt{4\pi}}\; . \label{stpt}
	\end{equation}
The nature of these points depends on the value of $\kappa$.
It is discussed below and in Appendix~B.

%
%
%
%

\section{Self-similar solutions}

The self-similar solutions fall quite naturally into two distinct classes
depending on the value of $\kappa$.  Class I solutions do not contain black
holes or naked singularities although they are singular at one
point on $r=0$.  In the
second class are solutions which have naked singularities or black
holes;  a
single critical evolution having a null singularity (which is {\em
not} naked)
interpolates between these two extremes.   For completeness, at the end of
this section it is shown how the
results of~\cite{Japs} fit into the overall picture.
%
%
%
%

\subsection{Class I -- $4\pi \kappa^2 >1$}

For $\kappa$ in this range only the stationary point $(y_+, {z}_+,
\gamma_+)$
is of interest. This exact solution is equivalently written as
	\begin{equation}
	g = 2 x (1 + \sqrt{4\pi}\kappa)\; , \ \
	\overline{g} =  x (1 + \sqrt{4\pi}\kappa) \; . \label{eqn:critpt1}
	\end{equation}
It has a singular origin ($r=0$),  in fact there are two sheets of this
singularity.  Future directed, ingoing lightrays terminate on the
sheet located at $u=0$,  while outgoing lightrays originate on the
past sheet (The solution may be obtained by setting $\alpha = \beta =
0$ in Eq.~(9) of \cite{Japs}).

Locally, (\ref{eqn:critpt1}) is a positive attractor (see appendix B).  The
global
structure  of the Class~I solutions is easily determined by
examining the behaviour of
the integral curves in the region
	\begin{equation}
	{\cal A} = \{1/(1 + 4\pi\kappa^2) < y < 1, \;  0<{z} <1/2\}\; .
	\end{equation}
Noting that integral curves enter ${\cal A}$ across the lines $y = 1/(1 +
4\pi\kappa^2)$,  $y=1$ and $\{ {z} =0, y \ge 1/2\}$ it is evident that the
solution originating at ${z} =0$,  $y=1$ either terminates at the stationary
point
or leaves ${\cal A}$ across ${z} = 1/2$.  In appendix C it is shown that
integral curves only cross ${z} = 1/2$ at $y = 1/(1 + 4\pi\kappa^2)$,  and
that the solution passing through this point is unique when $4\pi
\kappa^2 > 1$.  A direct consequence of
this is that the solution with a non-singular
 origin approaches the asymptotic form
(\ref{eqn:critpt1}), its evolution being characterised by a sequence of
decaying oscillations in $y$  about the value $y = 1/2$. (See Fig.~1)

\begin{figure}
\leavevmode
\hbox{\epsfxsize=16cm \epsfysize=16cm  {\epsffile{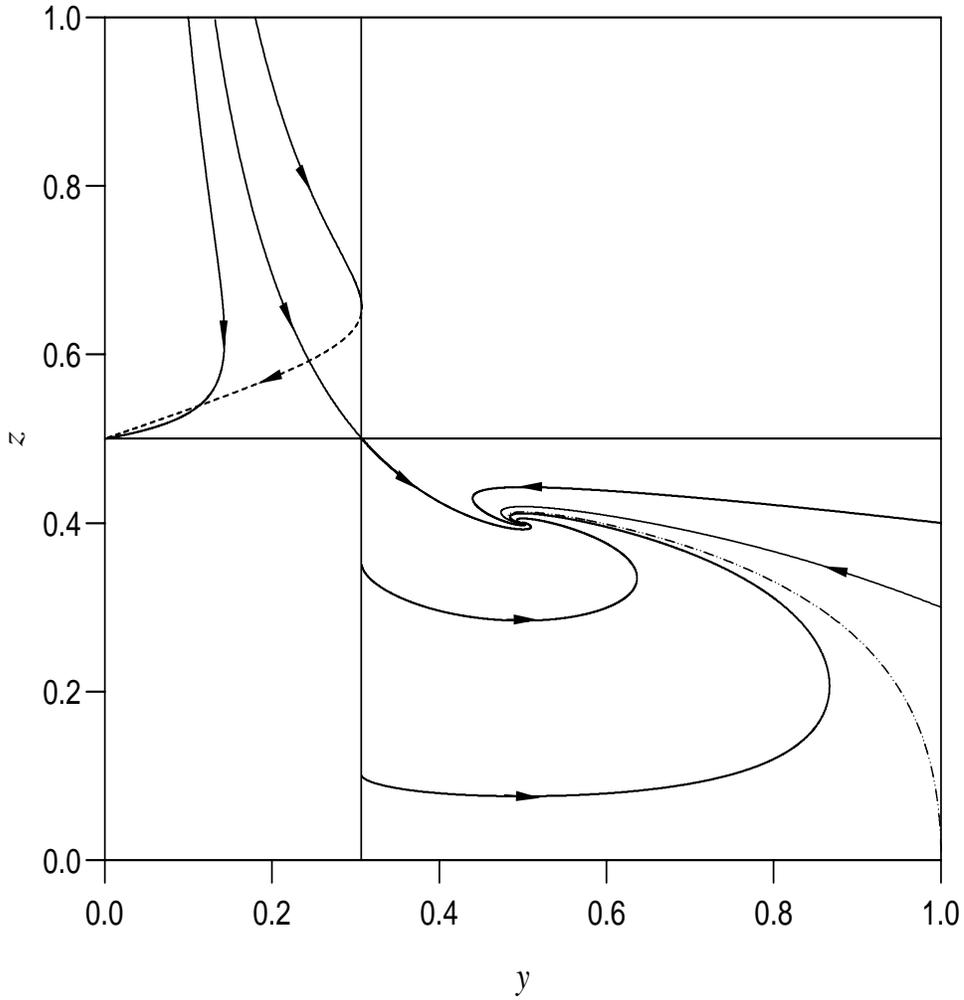}}}
\caption{ Integral curves when $4\pi\kappa^2 = 2.25$.  The solutions
are seen to spiral towards the stationary solution.  The dot-dashed
curve from $y=1$, ${z} =0$ is the solution with a non-singular
origin.  The dashed integral curve is on the negative leaf of
$\gamma$ and is the continuation of the solid curve which reaches $y = 1/(1 +
4\pi\kappa^2)$. }
\end{figure}

These solutions  do not contain black holes, yet there is a singularity at the
origin when $u\rightarrow 0$.  Is this singularity naked?  Consider ingoing,
radial null geodesics
	\begin{equation}
	\frac{d r}{du} = -  {\overline{g}}/2 \; .
	\end{equation}
Since $x = -r/u$ the geodesics are given by
	\begin{equation}
	\ln |u| = 2 \int^x_{x_1} (\overline{g} - 2x')^{-1} dx' \; ,
	\label{geo2}
	\end{equation}
where $x_1$ is the radius at the point of intersection between the ingoing
geodesic and the null cone $u=-1$.  Since ${z} = x/\overline{g} <1/2$
throughout the
entire
evolution the integrand in (\ref{geo2}) is always bounded,  and $x$ decreases
with increasing $u$.  Thus ingoing lightrays must reach $r=0$ before $u=0$,
and
they never reach the singularity provided $x_1 < \infty $.

In view of the asymptotic solution (\ref{eqn:critpt1}) one also finds that an
observer at a (large) fixed radius takes an infinite proper time to reach
$u=0$.
A central observer,  on the other hand, reaches the singularity at $u=0$
in finite proper time.  It is, however,  only after he has seen the
entire history of the universe in a tremendous flash.  Thus these spacetimes
have trivial topology,  with a singularity only at $r=0$ as $u\rightarrow 0$.

Similar arguments apply when $4\pi\kappa^2 =1$,  although in this case
$\overline{g} -
2x \rightarrow 0$  as $x\rightarrow \infty$.
It must be emphasised that this solution is unstable,  in the
sense that an arbitrarily small change in the value of $\kappa$
drastically changes the character of the resulting solution.  In
particular,  for smaller values the spacetime contains a black hole.

%
%
%
%

\subsection{Class II -- $4\pi\kappa^2 <1$}

Containing two
sub-classes of solutions and offering
another example of a phase transition in gravitational collapse -- from black
hole spacetimes to naked singularities -- these solutions are more interesting
than those in Class~I.  In fact for each $\kappa$ in the range
$0 < 4\pi\kappa^2 <1$
there appears to be a continuous infinity of solutions with a regular origin.
This is due to the
failure of uniqueness at the singular point (\ref{eqn:singpt}).  Naked
singularities develop only for sufficiently small values of $\kappa$.

Solutions with a non-singular origin contain a null hypersurface, $\Gamma$ say,
on which $x =$~constant.  This corresponds to the point $y=
1/(1+4\pi\kappa^2)$,
${z} = 1/2$ in phase space.  As mentioned earlier the standard uniqueness
theorems break down at this point,  and there is a one parameter family of
self-similar continuations beyond it.  More general extensions which produce
asymptotically flat
spacetimes have been considered in~\cite{Goldpir}.

\begin{figure}
\leavevmode
\hbox{\epsfxsize=16cm \epsfysize=16cm  {\epsffile{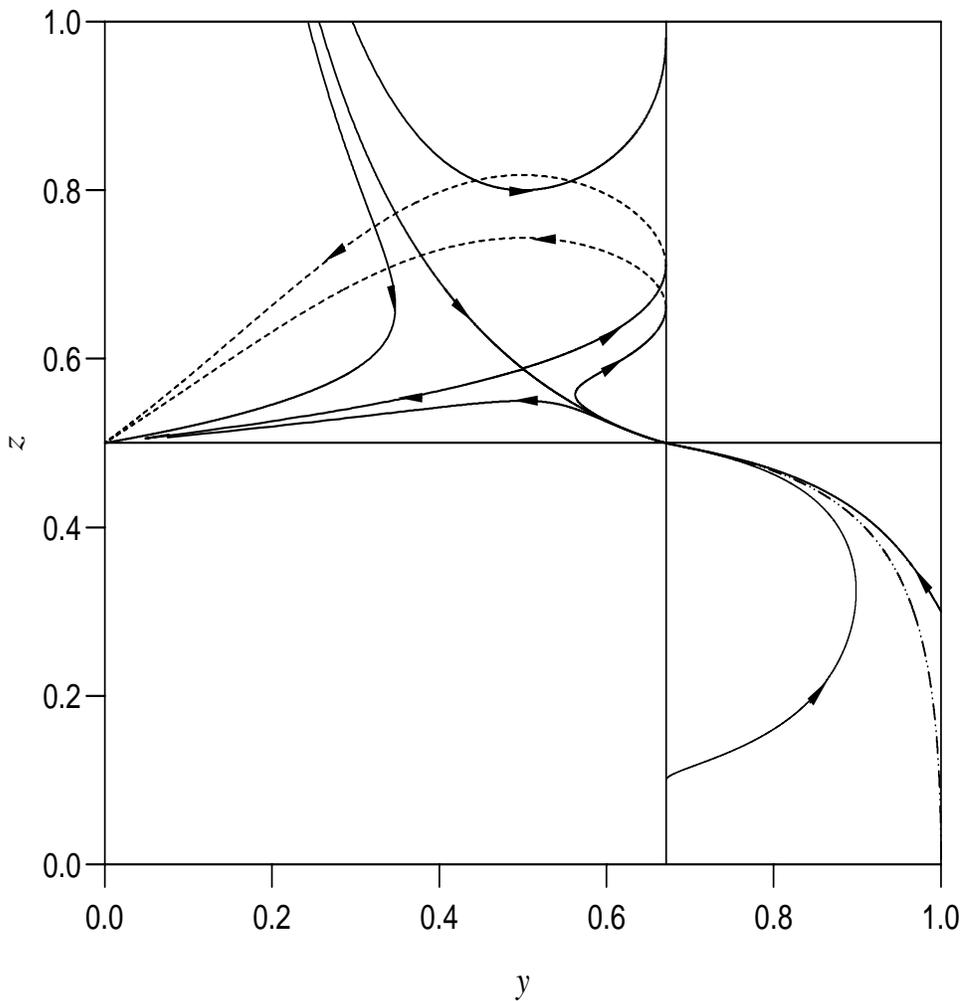}}}
\caption{Integral curves when $4\pi\kappa^2 = 0\cdot 49$.  Here we
see the integral curve representing the solution with a non-singular
origin reach the point ${z} = 1/2$, $y = 1/(1 + 4\pi\kappa^2)$.
There is a one parameter family of continuations past this point. All
solutions of interest terminate at $y=0$ (an apparent horizon in
spacetime) except one
which approaches the stationary point.  Here also the
dashed lines are integral curves on the
negative leaf of $\gamma$.}
\end{figure}

\begin{figure}
\leavevmode
\hbox{\epsfxsize=16cm \epsfysize=16cm  {\epsffile{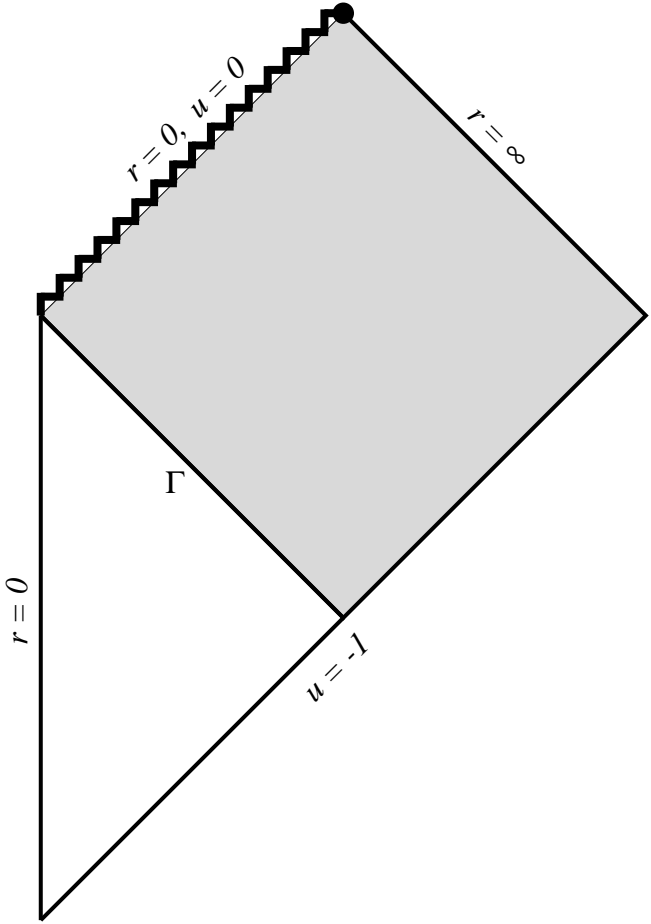}}}
\caption{A spacetime diagram for the critical evolution when
$4\pi\kappa^2 <1$.  The unshaded region corresponds to the
dot-dashed curve in Fig.~2 and Fig.~4.  The shaded region shows a continuation
past $\Gamma$ [where ${z} = 1/2$, $y = 1/(1 + 4\pi\kappa^2)$] which
approaches a stationary point.  The singularity at $r=0$ is null.}
\end{figure}

Fig.~2 shows various solutions to the system
(\ref{eqn:aut1})-(\ref{eqn:aut3}) when $4\pi \kappa^2 = 0\cdot 49$.  It is
clear
from
the diagram that all solutions (except one) evolve into black holes (signalled
by the formation of an apparent horizon,  $y\rightarrow 0$).  Some of
the solutions evolve onto the negative leaf of (\ref{gamma}); $y$ undergoes a
single oscillation before decreasing  monotonically to $y=0$.  While I have no
analytic proof that all of these solutions contain black holes,  the
numerical results (as illustrated in Fig.~2) suggest
this is true.  The single exceptional solution exhibits a behaviour which has
been discussed elsewhere~\cite{Japs};  beyond $\Gamma$ no black hole forms,
instead
the solution asymptotically approaches one of the stationary points
(\ref{stpt}).  The
spacetime is singular at the null surface $u=0$, $r=0$ where $y=1/2$. This
singularity lies at infinite redshift for observers at large radius.  Fig.~3
is a spacetime diagram for this
exceptional case.

\begin{figure}
\leavevmode
\hbox{\epsfxsize=16cm \epsfysize=16cm  {\epsffile{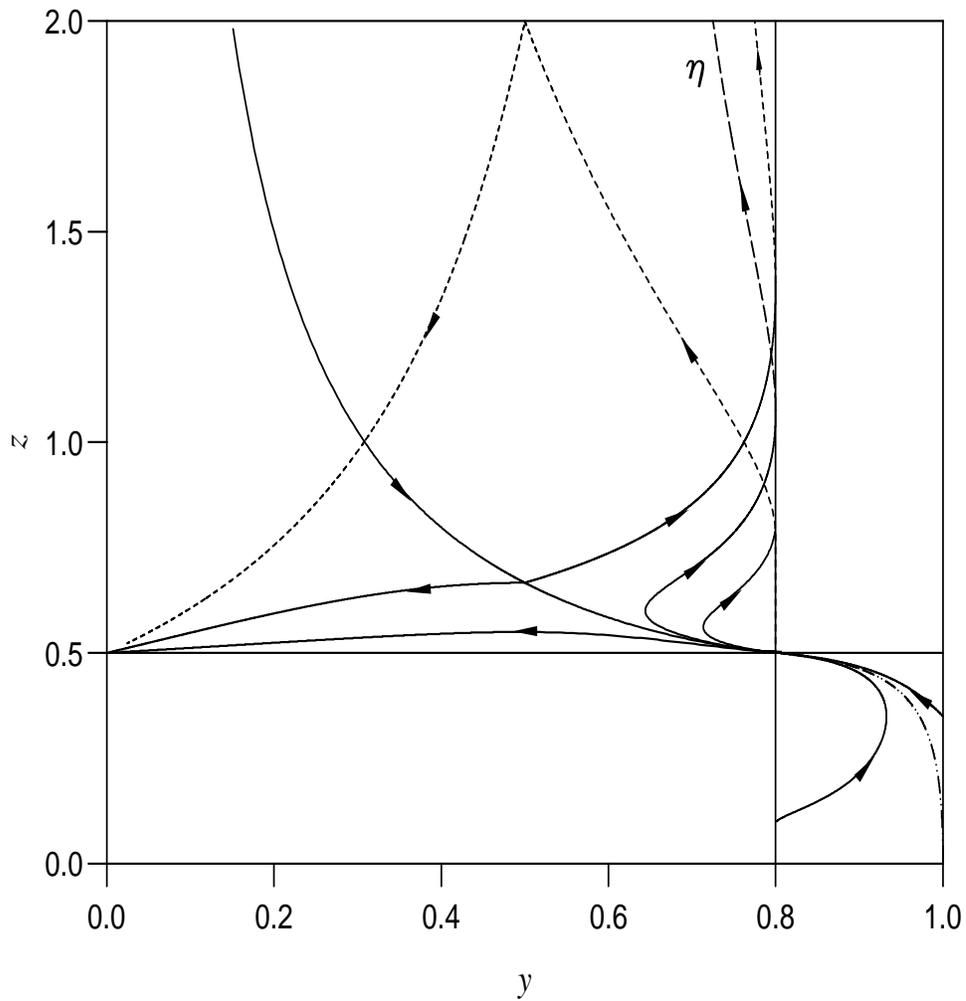}}}
\caption{Integral curves when $4\pi\kappa^2 = 0\cdot 25$.  $\eta$ is an example
of an integral curves which evolves onto the
negative leaf of $\gamma$ but does not reach $y=0$,  instead it attains a
minimum value of $y$ and then asymptotes to $y = 1/(1 + 4\pi\kappa^2)$. Such
solutions have a naked singularity at $r=0$.}
\end{figure}

For smaller values of $\kappa$ more complicated behaviour is possible,  and
indeed another type of phase transition is apparent -- from black holes to
naked
singularities.  When ${z} >1/2$ the $x =$ constant surfaces are spacelike.
If no black hole
forms  then $x\rightarrow \infty$ as $u\rightarrow 0$ [see Eq.~(\ref{geo2})],
corresponding to a Cauchy horizon in the spacetime.  For sufficiently small
$\kappa$ there are solutions which evolve from the singular point onto the
negative leaf of (\ref{gamma}), asymptoting to $y = 1/(1 + 4\pi\kappa^2)$ as
${z} \rightarrow \infty$. A typical solution of this type is shown in
Fig.~4,  for $4\pi\kappa^2 = 0\cdot 25$.  It is not difficult to obtain an
approximate solution in the large ${z}$ limit,  and hence to show that the
Cauchy horizon is non-singular.  For large ${z}$ Eq.~(\ref{eqn:aut2}) becomes
	\begin{equation}
	\dot{y} \simeq 1 - (1+4\pi\kappa^2)y \; .
	\end{equation}
Integrating and inserting the result in (\ref{eqn:aut1}) implies
	\begin{eqnarray}
	y &\simeq& \frac{1}{1+4\pi\kappa^2} + c x^{-(1+4\pi\kappa^2)} \\
	{z} &=&  \frac{x}{\overline{g}} \simeq d x ^{1-4\pi\kappa^2} \; ,
	\end{eqnarray}
where $c$ and $d$ are arbitrary constants of integration.
It is now straightforward to verify the regularity of the Ricci scalar on the
Cauchy horizon,  using (\ref{gamma}) and (\ref{eqn:T})
	\begin{equation}
	\left. T{^{\alpha}}_{\alpha} \right|_{u=0} = \frac{\kappa^2}{\Delta
	r^2} \; .
	\end{equation}
Only as $r\rightarrow 0$ is this quantity singular,  indicating the existence
of
a  naked singularity at $u=0$, $r=0$ in these solutions.  The Cauchy horizon is
also a null
orbit of the homothetic Killing vector,  therefore the work of Lake
and Zannias~\cite{L&Z}
implies that the singularity is strong in  the sense that tidal forces diverge
at it.
Notice also that  the metric can be made
manifestly regular at the Cauchy horizon  by transforming to the new
coordinate given by
	\begin{equation}
	dU = (-u)^{-4\pi\kappa^2} du\; .
	\end{equation}
These  spacetimes confirm that naked singularities, which evolve from regular
initial data, also exist for scalar field sources.
Whether they are stable to non-homothethic,  not to mention
non-spherical,  perturbations is an open question although the work of
Goldwirth
and Piran~\cite{Goldpir} suggests that they are not.

There are two critical points where phase transitions are observed.   Each of
the exactly critical solutions, which interpolates between naked singularities
and black holes,  asymptotically approaches one of the stationary
points~(\ref{stpt}). Therefore these spacetimes have the structure shown in
Fig.~3, with  a null singularity at $u=0$, $r=0$.

As $\kappa$ decreases further,  the topology of the phase space changes
slightly;  only a single phase transition occurs going from black holes to
naked
singularities.  The solutions (including the single critical evolution) are the
same as those already discussed.

%
%
%
%

\subsection{The Roberts solution -- $\kappa =0$}

\begin{figure}
\leavevmode\centerline{
\hbox{\epsfxsize=16cm \epsfysize=16cm  {\epsffile{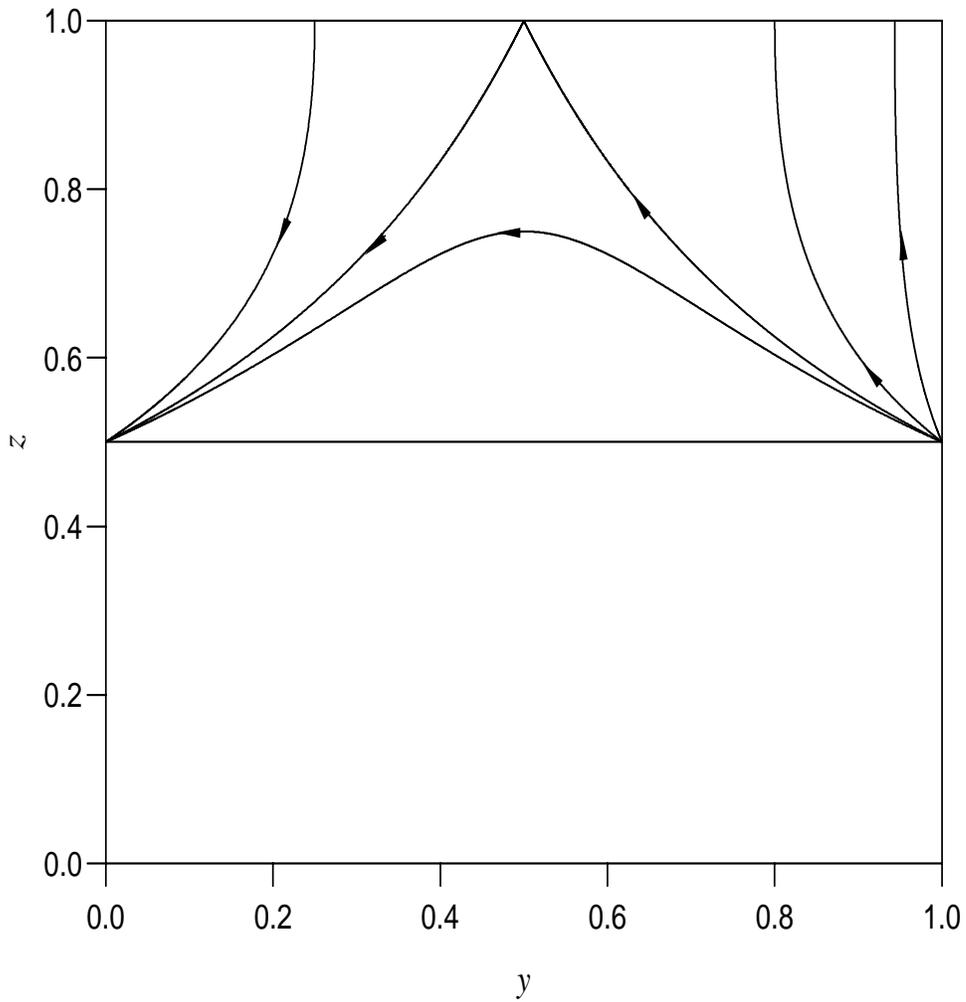}}}}
\caption{The exact
expression for the solutions when $\kappa =0$ is given in the text,  however
this diagram is for comparison with the other cases.  Note that y=1 is
now a solution to the equations and  corresponds to flat space.  For
this reason no naked singularities exist for any value of $p$ when $\kappa =
0$.}
\end{figure}

When $\kappa =0$ the scalar field is a function of $x = -r/u$ only and the
equations (\ref{eqn:ss1})-(\ref{eqn:ss4}) can be exactly
integrated~\cite{Japs}.
 The resulting solution was first discovered by Roberts~\cite{Roberts} and may
be written as
	\begin{eqnarray}
	\overline{g} &=& \{( p^2 + x^2 )^{1/2} - 2 p^2\}/x\\
        g&=& 2\overline{g} \left( 1 \pm \sqrt{1 - 4p^2(2x\overline{g} -
\overline{g}^2)}\,
            \right)^{-1}
	\end{eqnarray}
and
	\begin{equation}
	\gamma = p/\sqrt{4\pi  ( p^2 + x^2)}\; .
	\end{equation}
The integral curves are plotted in  Fig.~5.  The solutions are labelled by $p$
and
exhibit critical point behaviour.  When $p = 1/2$ the integral curve leaves
${z} = 1/2$, $y=1$ and approaches the stationary point (\ref{stpt}).  This
exactly critical evolution lies between solutions which contain black holes
($y\rightarrow 0$, ${z} \rightarrow 1/2$ along the integral curves)  and
those
which evolve back to flat space ($y\rightarrow 1/(1 + 4\pi\kappa^2)$, ${z}
\rightarrow
\infty$ along the integral curves).  Since this is a saddle point we
see that near critical evolutions (sub- or super-critical) can
approach this point arbitrarily closely before moving away in their
respective directions.

$\kappa =0$ is therefore exceptional because the subcritical evolutions have
zero mass on the Cauchy horizon, $u=0$,  as is readily seen by taking the limit
$x\rightarrow \infty$ in the above solution.  This behaviour is also apparent
in
the asymptotic approach of the integral curves to the line $y=1$.  As discussed
in~\cite{Japs} the natural continuation past $u=0$ is Minkowski space since
there is no material flux across this surface on which $m=0$.
Notice however that there is no self-similar extension of these solutions to
$r=0$ which is non-singular.

This discussion places the results obtained in \cite{Japs} within the more
general context of self-similar spacetimes with scalar field matter sources.

%
%
%
%

\section{Discussion}

Spherically symmetric,  homothetic spacetimes have received a great deal of
attention over the last few years due to the ease with which it is possible to
construct naked singularities in such spacetimes.  Recent numerical studies of
spherical collapse~\cite{Chop,Evans} suggest that self-similarity may play an
important
role in describing the approach to the singularity in gravitational collapse.
This study of scalar field collapse was in fact motivated by the work of
Choptuik,  where he observed discrete self-similarity in solutions on the verge
of black hole formation.  However it is not clear how a continuous self
similarity, as discussed here, could be at the center of the results which he
has
obtained.    Nevertheless some interesting features do emerge from the study of
spacetimes with a homothetic symmetry.

The scalar field evolves as  $\psi = \overline{h} (r/|u|) - \kappa \ln |u|$
where
$\overline{h} (r/|u|)$ is determined by the coupled Einstein-scalar field
equations.
The constant $\kappa$ distinguishes between two different classes of
solution.
The first class (when $4\pi\kappa^2 \ge 1$) are non-singular almost everywhere.
Technically there is a singularity at $u = 0 = r$,  however it is not naked and
the worldlines which reach it in finite proper time are a set of
measure zero.
When $0 \le 4\pi\kappa^2 <1$  the solutions are more interesting --
for each value of $\kappa$ in this range there exists infinitely many solutions
with a non-singular origin.  When $4\pi\kappa^2$ is only slightly
less than
unity all of these spacetimes contain apparent horizons.  However
for {\em sufficiently} small $\kappa$ some of the spacetimes have naked
singularities.

 $4\pi \kappa^2 =1$ marks the  transition point from solutions of Class~I to
those of Class~II -- in effect from no black holes to black holes.  Another
phase transition occurs for each value of $\kappa < 1/\sqrt{4\pi}$ -- a
transition from black holes to naked singularities.
In particular the critical evolution is on the verge of being a
naked singularity in this case.  One might wonder if there is some
way to classify the critical point behaviour more generally,  and to
which such class the {\em generic} case belongs.

One of the most intriguing results obtained by Choptuik was the scaling law for
black hole mass, unfortunately self-similar spacetimes cannot have finite mass
black holes.  In order to obtain an asymptotically flat spacetime it is
necessary to cut off the self-similar evolution at some advanced time,  and
consider a suitable continuation (which is not self-similar).  Goldwirth and
Piran~\cite{Goldpir} did exactly this,  although they did not examine the
behaviour of
black hole mass as the critical point $4\pi\kappa^2$ was approached.  This
question is currently under active investigation~\cite{John},  and it will be
interesting to see if the mass exhibits the same behaviour which has been
observed elsewhere~\cite{Chop,Evans,Abrahams}.

We have in self-similar scalar field collapse further
examples of spacetimes which violate cosmic censorship.  One might
be surprised about this were it not for the
plethora of examples which now exist.  What emerges from these examples
(generally) is that there do exist initial data sets which when evolved
according to the Einstein equations lead to naked singularities,  however the
genericity of these data is far from clear.  It therefore seems that the thrust
of any attempt to formulate (and prove) cosmic censorship must address this
issue directly.  Some interesting preliminary results have been obtained by
Lake~\cite{Lake} where he has shown that (spherically symmetric) spacetime in
the
neighbourhood of a naked singularity may be approximately self-similar.
It therefore seems that future work on naked singularities must consider
deviations from the symmetric situations treated to date.

Finally in searching for a theoretical understanding of the results
obtained by Choptuik~\cite{Chop}  and Abrahams and Evans~\cite{Abrahams} one
might consider the obvious generalisation of hometheticity to a
conformal symmetry;  that is to suppose the existence of a vector
field $\xi$ such that
    \begin{equation}
    {\cal L}_{\xi} g = \Omega(x) g \; .
    \end{equation}
If the dependence on position in $\Omega$ is weak,  solutions might
behave like self-similar solutions with some sort of super-imposed
periodicity.  It would seem interesting to investigate this
possibility.

\section*{Acknowledgements}

I would like to thank Chris Chambers, Peter Hogan, Bruce Jensen, Ian
Moss and Adrian Ottewill for useful comments on this work.  It is
also  a pleasure to thank Carsten
Gundlach  whose comments and questions on an earlier draft of this paper
have lead to a significantly different final version.
This work was supported by EPSRC of Great Britain.

%
%
%
%

\appendix
\section{Scalar field evolution in spherical, self-similar spacetimes}

Results of Defrise-Carter~\cite{Defrise-Carter} imply that a spherical
spacetime
with a homothetic symmetry (i.e. there exists a vector $\xi$ such that ${\cal
L}_{\xi} g = 2 g$) can be written in the form
	\begin{equation}
	ds^2 = e^{2t} \left( g_1(x) dt^2 + g_2(x) dx^2 + e^{2x} d\Omega^2\right)
	\label{A1}
	\end{equation}
where $d\Omega^2 = d\theta^2 + sin^2 \theta \, d\phi^2$,  and the similarity
vector is $\xi = \partial / \partial t$.

The exponential dependence of the metric on $t$ gaurantees that the Christoffel
symbols and hence the Ricci tensor,  $R_{\mu\nu}$, are independent of this
coordinate.  This may be expressed covariantly as
	\begin{equation}
	{\cal L}_{\xi} R_{\mu\nu} = 0
	\; . \label{A2}
	\end{equation}
Before examining the implication of this for a self-similar spacetime which
satisfies Einstein's equations with scalar field matter,  let me show that
(\ref{A1}) can be recast into the form used in section~II.

Introduce new coordinates $r$ and $u$ defined by
	\begin{equation}
	r = \exp({t + x}),  \ \ \ u = r G(x), \label{A4}
	\end{equation}
where  $G(x)$ is to be determined.  Substituting them into the line element
(\ref{A1}) and requiring $u$ to be null one obtains the ordinary differential
equation
	\begin{equation}
	\frac{dG}{dx} = G \left( - 1 \pm \sqrt{{-g_2(x)}/{g_1(x)}}\right) \;
	\end{equation}
which determines $G(x)$.  Furthermore the line element reduces to
	\begin{equation}
	ds^2 = -\overline{g}(r/u)g(r/u) du^2 - 2g(r/u) dudr + r^2 d\Omega^2 \; ,
	\end{equation}
where $r/u$ is realted to $x$ by (\ref{A4}).  Clearly this means that $g$ and
$\overline{g}$ are
functions of $r/u$ as stated in section~II.  The similarity vector is
	\begin{equation}
	\xi = r \partial_r + u\partial_u \label{A7}
	\end{equation}
in these coordinates.

Now in view of (\ref{A2}) and the Einstein field equations
	\begin{equation}
	R_{\mu\nu} = 8\pi \psi_{,\mu} \psi_{,\nu}
	\end{equation}
the scalar field must satisfy
	\begin{equation}
	{\cal L}_{\xi} (\psi_{,\mu}) = 0 \; . \label{A9}
	\end{equation}
Assuming that $\psi$ is independent of $\theta$ and $\phi$,  Eq.~(\ref{A9}) is
readily integrated to
	\begin{eqnarray}
	\frac{\partial \psi}{\partial r} &=& \frac{\gamma(r/u)}{r} \; \\
 	\frac{\partial \psi}{\partial u} &=& \frac{-\gamma(r/u)}{u}\;
 	\end{eqnarray}
where $\gamma$ is an arbitrary function of $r/u$.  (Integrability was used to
reduce the number of arbitrary functions to one)  The general solution of these
coupled equations is therefore
	\begin{equation}
	\psi = \overline{h}(r/u) - \kappa \ln|u| - \beta \ln|r| \; .
	\label{A11}
	\end{equation}
In particular $\beta$ may be set to zero by absorbing $\beta(\ln|u| - \ln|r|)$
into $\overline{h}$,  thus reducing (\ref{A11}) to (\ref{eqn:sf}).

\section{The stationary points}

The discussion in section III relies on the properties of the stationary points
of the equations (\ref{eqn:aut1})-(\ref{eqn:aut3}).  In general there exists
two
such points given by
	\begin{equation}
	y_{\pm} = \frac{1}{2}\; , \ \  {z}_{\pm} = \frac{1}{1 \pm \sqrt{4\pi}
	\kappa} \; , \ \
	\gamma_{\pm} = \pm \frac{1}{\sqrt{4\pi}}\; .
	\end{equation}

	Provided $4\pi\kappa^2 \ne 1$ it is straightforward to linearise about
each of these points and hence to analyse the topology of the phase space in
their neighbourhoods.  The eigenvalues are
	\begin{equation}
	\lambda_{1,2} = \frac{1-{z}_{\pm}}{2{z}_{\pm} -1}
	\pm \sqrt{ \frac{ (1 - {z}_{\pm})^2 + 4 (2{z}_{\pm} -1)}{
	(2{z}_{\pm} -1)^2}} \; ,
	\end{equation}
where ${z}_{\pm}$ may be chosen independently of the sign of the square
root.

When $4\pi\kappa^2 >1$  only ${z}_+$ is relevant for the discussion in
section III.  Both eigenvalues are real and have the same sign, $\lambda_{1,2}
<0$, when $1 < 4\pi\kappa^2 \le 4/3$  so that the stationary point is an
attractive node.  Once  $4\pi\kappa^2 > 4/3$ the eigenvalues become complex
conjugate,  and since
	\begin{equation}
	2{z}_{+} -1 <0\; , \ \  1 - {z}_+ >0
	\end{equation}
they have negative real part.  This is a positive attractor,  with spiral
behaviour.

When $0 \le 4\pi\kappa^2 <1$ both
stationary points are of interest.  Simply noting that
	\begin{equation}
	2{z}_{\pm} -1 > 0\; , \ \  1 - {z}_+ >0 \; , \ \  1 - {z}_- <0
	\end{equation}
the discussion of both is easily combined.  Clearly  the eigenvalues are real
and have opposite signs since
	\begin{equation}
	|1 - {z}_\pm | < \sqrt{ (1 - {z}_{\pm})^2 + 4 (2{z}_{\pm} -1)}\; .
	\end{equation}
Thus they are saddle points.

\section{The singular line ${z} = 1/2$}

In the above analysis it is important that integral curves cannot cross ${z}
=
1/2$ except at $y=0$ or at $y = 1/(1 + 4\pi\kappa^2)$.  We now show that this
is so,  and
derive the solution in the neighbourhood of ${z} = 1/2$, $y=1/(1 +
4\pi\kappa^2)$.  The
analysis is split into two cases:
\\
\mbox{\ }
\\
(i) Suppose an integral curve crosses ${z} = 1/2$ at $y_0 \ne  1/(1 +
4\pi\kappa^2)$.
Writing ${z} = 1/2 + \zeta$ and considering the $\zeta \rightarrow 0$ limit
of
(\ref{eqn:aut1}), (\ref{eqn:aut2}) and (\ref{gamma}) it is a straightforward
matter to derive
	\begin{equation}
	\frac{dy}{d\zeta} \simeq \frac{(1 + 4\pi\kappa^2) y^2 - y}{(2y - 1)
	\zeta} \;
        {}.
	\end{equation}
Integrating this equation gives
	\begin{equation}
	\ln |\zeta| \simeq   \mbox{constant} +
			\frac{(1 - 4\pi\kappa^2)}{(1 + 4\pi\kappa^2)}
			\ln  |(1 + 4\pi\kappa^2) y^2 - y |  \; .
	\end{equation}
Examining this expression shows that $y_0 =0$ is the only place where integral
curves may intersect ${z} = 1/2$.
\\
\mbox{\ }
\\
(ii) It is necessary to treat the case when $\left. \zeta\right|_{y=1/(1 +
4\pi\kappa^2)} =
0$ separately since the limit ${z} \rightarrow 1/2$ in (\ref{gamma}) is more
delicate.  For this purpose we introduce $\zeta$ as above and write
	\begin{equation}
	y = \frac{1}{1 + 4\pi\kappa^2} + \eta
	\end{equation}
where $\eta \ll 1/(1 + 4\pi\kappa^2)$,  thus
	\begin{equation}
	\gamma \simeq -\kappa \pm
\frac{1 + 4\pi\kappa^2}{\sqrt{8\pi}}\sqrt{\frac{\eta}{-\zeta}}\; .
	\end{equation}
Substituting this approximate expression for $\gamma$ into (\ref{eqn:aut2}),
and using
$\zeta$ as the independent variable we arrive at the equation
	\begin{equation}
	\frac{d\eta^{1/2}}{d\zeta} - \frac{1 + 4\pi\kappa^2}{2\zeta (1 -
4\pi\kappa^2)}
	\eta^{1/2} \simeq \pm \frac{\kappa \sqrt{8\pi}}{(-\zeta)^{1/2}
	(1 - 4\pi\kappa^2) }
	\end{equation}
and find
	\begin{equation}
	\eta^{1/2} \simeq \frac{\sqrt{8\pi}}{4\pi \kappa} (-\zeta)^{1/2}
	+ {c}  (-\zeta)^{(1 + 4\pi\kappa^2)/(1 - 4\pi\kappa^2)} \;  ,
	\end{equation}
where $c$ is a constant of integration.
Now the initial condition is $\eta = 0$ when $\zeta  =0 $ so that there are two
distinct possibilities;  when $4\pi\kappa^2 >1$ the integration constant must
vanish,  implying that a single integral curve passes through this point.  On
the other hand if $4\pi\kappa^2 <1$  the constant is not fixed by the initial
conditions and there is a one parameter family of curves passing through $\eta
=
0 = \zeta$.  It is exactly this fact which gives rise to the variety of
solutions in  Class~II.

\end{document}